\documentclass[11pt,a4paper]{article}
\pdfoutput=1
\usepackage{empheq}% http://ctan.org/pkg/empheq (autoloads amsmath)
\usepackage{cases}% http://ctan.org/pkg/cases
\usepackage{a4wide}
\usepackage{amsfonts}
\usepackage{amssymb}
\usepackage{amsmath,bm}
\usepackage{amsmath}
\usepackage{graphicx}
\usepackage{mathtools}
\usepackage{booktabs}
\usepackage{array}
\usepackage{color}
\usepackage{ulem}
\usepackage[small,bf]{caption}
\usepackage{cite}
\usepackage[usenames,dvipsnames]{xcolor}
\usepackage{subfigure}
\usepackage{verbatim}
\usepackage{xcolor}
\allowdisplaybreaks

\numberwithin{equation}{section}

\newcounter{mysubequation}[equation]

\DeclarePairedDelimiter\bra{\langle}{\rvert}
\DeclarePairedDelimiter\ket{\lvert}{\rangle}
\DeclarePairedDelimiterX\braket[2]{\langle}{\rangle}{#1 \delimsize\vert #2}
\begin{document}
\begin{titlepage}

\begin{center}
{
\bf\LARGE Discovery Potentials for the Cosmic Neutrino Background\\[0.3em]
using Single Arm Interferometer
}
\\[8mm]
Chrisna~Setyo~Nugroho$^{\, a}$ \footnote[1]{chrisna@apps.ipb.ac.id, setyo13nugros@gmail.com}
\\[1mm]
\end{center}
\vspace*{0.50cm}

\vspace*{0.50cm}
{\centering \it
$^{a}$ Theoretical Physics Division, Department of Physics, IPB University,\\Jl. Meranti Wing S Level 5,
Kampus IPB Dramaga, Bogor 16680, Indonesia\\
}
\vspace*{1.20cm}

%\centerline{\it Department of Physics, National Taiwan Normal University, Taipei 116, Taiwan}
%\centerline{\it Theoretical Physics Division, Department of Physics, IPB University,}
%\centerline{\it Jl. Meranti Wing S Level 5,
%Kampus IPB Dramaga, Bogor 16680, Indonesia}
%\vspace*{1.20cm}

\begin{abstract}
\noindent
We study the interaction between light and non-relativistic cosmic neutrino background (CNB). We propose
single arm laser interferometry with coherent laser source and two distinct squeezing operators to
probe such interaction. We analyze the induced phase shift in four distinct quantum regimes of the interferometer operation:
the standard quantum limit (SQL), the Heisenberg limit, super-Heisenberg limit with $N^{3/2}$ enhancement, and
super-Heisenberg limit with $N^{2}$ enhancement where $N$ is the number of photons in the
interferometer. We demonstrate that the $N^{2}$ super-Heisenberg enhancement has the potential to probe the
magnetic moment down to $10^{-25}\, \mu_{B}$ where $ \mu_{B}$ is the Bohr magneton. If the CNB and 
photons interact via its magnetic moment, this interaction could be revealed in single arm interferometer operating at super-Heisenberg limit with $N^{2}$ enhancement.
\end{abstract}

\end{titlepage}
\setcounter{footnote}{0}

\section{Introduction}

As a remnant of the Big Bang, relic neutrinos known as cosmic neutrino background (CNB) permeate
the universe. The detection of such relic would deepen our understanding of the early universe as well
as the nature of the neutrino itself. However, the current CNB temperature is $\text{T}_\nu = 1.95 \; \text{K}$ which corresponds to the mean momentum value $\langle p_\nu \rangle = 0.529$ meV. This very low
energy as well as its weak interaction nature are the main obstacle of the CNB detection. 
Several detection methods have been proposed to probe the CNB. In PTOLEMY experiment, they expect to observe
the CNB capture on tritium targets~\cite{PTOLEMY:2018jst,Betts:2013uya,Banerjee:2023lrk}. The
scattering of the CNB with the detector of the gravitational wave experiments has also been considered~\cite{Domcke:2017aqj,Shergold:2021evs,Nugroho:2025ufk,Nugroho:2023cun,Chen:2025tlx}. Furthermore, high
energy cosmic neutrino and CNB scattering would leave a detectable effect as discussed in ~\cite{Brdar:2022kpu,Mohammadi:2021xoh,Asteriadis:2022zmo}. Moreover, there are many other detection methods that have been proposed to detect the CNB ranging from laboratory experiments as well as astrophysical observations~\cite{Bauer:2023CNBlimits,Bernal:2021RadiativeCNB,Yoshimura:2015RelicNu,Bauer:2021RelicAcc,Brdar:2022CosmogenicProbe,Franklin:2025TeVCNB,Chauhan:2025NSprobeCNB, Das:2025OldNSRelic,Bauer:2025DMPairAbsorption,Stodolsky:1975NeutrinoSea,Rostagni:2023DarkStodolsky}.
 
On the other hand, recent progress on quantum technology has entered a new era where the quantum
phenomena have been observed  and even manipulated directly in laboratories. Its well known application
in fundamental science was realized in the detection of gravitational wave (GW) at
LIGO~\cite{LIGOScientific:2016aoc}. There, the quantum nature of the light known as photon was utilized
in the interferometry setup to detect the phase shift induced by gravitational wave. The standard
quantum limit (SQL) has been achieved in LIGO which corresponds to the sensitivity of the phase shift
measurement as tiny as $\delta \geq 1/\sqrt{N}$ where $N$ is the number of
photons inside the interferometer, typically of the order of $\mathcal{O}(10^{23})$ or
larger~\cite{aLIGO:2016pgl}. Furthermore, the Heisenberg limit corresponding to the phases shift measurement $\delta \geq 1/N$ has been achieved as well~\cite{Szigeti:2017,Linnemann:2016,Daryanoosh:2018,Anderson:2017}. This limit is originated from quantum mechanics which gives the best sensitivity one could reach in a quantum measurement~\cite{Dirac:1927,Heitler:1954}. However, this is not the end of the story as the Heisenberg limit has been surpassed in the experiment using entangled photon source with the sensitivity reach $\delta \geq 1/N^{3/2}$ known as super-Heisenberg limit~\cite{Napolitano:2011,Napolitano:2011b} (super-Heisenberg with $N^{3/2}$ enhancement where $N$ is the number of the probe). In addition, quantum protocol implemented in quantum optical experiment has been shown to achieve better super-Heisenberg sensitivity with $\delta \geq 1/N^{2}$~\cite{Qin:2023ift,Hou:2021qnd}.         

The impressive sensitivity reach of the current quantum optical experiments, especially in the
interferometry setup, has motivated us to consider the possibility to detect the CNB using laser interferometer.      
However, one should be aware that the available exotic particle search at Gravitational Wave (GW) experiments~\cite{Tsuchida:2019hhc,Lee:2020dcd,Chen:2021apc,Ismail:2022ukp,Lee:2022tsw,Seto:2004zu,Adams:2004pk,Nugroho:2024ltb,Riedel:2012ur,PhysRevLett.114.161301,Arvanitaki:2015iga,Stadnik:2015xbn,Branca:2016rez,Riedel:2016acj,Hall:2016usm,Jung:2017flg,Pierce:2018xmy,Morisaki:2018htj,Chen:2022abz,Grote:2019uvn} can not be utilized to detect the CNB since both arms of the interferometer experience the
uniform CNB background leading to zero phase shift at the detection port. To solve this issue, we
propose to utilize single arm laser interferometer equipped with two spatially separated squeezing and
anti-squeezing devices to detect the CNB. This interferometer is originally proposed by~\cite{Yurke:1986} and has been widely used in laboratories~\cite{Hudelist:2014,Manceau:2016esq,Liu:2018ahw,Xiao:2019,Ferreri:2020pxa}. Moreover, the application of this kind of interferometer to search for dark matter has been proposed recently~\cite{Capolupo:2025zei}.

The remainder of this paper is structured as follows. In Section~\ref{sec:interaction}, we discuss the
interaction between non-relativistic Dirac neutrinos and photons. Section~\ref{sec:phase} explains a
particular phase measurement scheme in single arm interferometer to probe CNB-photons interaction, and
in Section~\ref{sec:result}, we evaluate the corresponding projected sensitivities of the proposed setup. Finally, Section~\ref{sec:Summary} provides a summary as well as conclusion of our results.

%Also, we fix the neutrino number density in the universe, $n_{\text{CNB}}$, equals to $56 \, \text{cm}^{-3}$ per degree of freedom~\cite{Ringwald:2004np}. 

\section{CNB and Photon Interaction}
\label{sec:interaction}

We consider massive Dirac neutrinos with magnetic moment $\mu^{\text{D}}_{kj}$ where the indices $k,j$
denote the mass basis. The Lagrangian density describing neutrino electromagnetic interaction via magnetic moment is given by~\cite{Giunti:2024gec} 
\begin{align}
\label{eq:LDirac}
\mathcal{L}^{\text{D}}_{\text{mag}} =  -\frac{1}{2}\, \sum_{k,j}\,\bar{\nu}_{k}\,\sigma^{\alpha \beta}\,\mu^{\text{D}}_{kj}\, \nu_{j}\,F^{\alpha \beta}\,,
\end{align}
where $\nu_{k}$ and  $F_{\alpha \beta} = \partial_{\alpha} A_{\beta}- \partial_{\beta} A_{\alpha}$ stand for the neutrino field and 
electromagnetic tensor, respectively. Note that the explicit expression of the magnetic moment $\mu^{\text{D}}_{kj}$
is model dependent. In the minimal extension of the Standard Model (SM) with three massive Dirac
neutrinos, the magnetic moment can be written as~\cite{Fujikawa:1980yx,Pal:1981rm,Shrock:1982sc,Dvornikov:2003js,Dvornikov:2004sj,Giunti:2014ixa}  
\begin{align}
\label{eq:mmoment}
\mu^{\text{D}}_{kj} \simeq \frac{3 \,e \,G_{\text{F}}}{16\,\sqrt{2}\,\pi^{2}}\,(m_{k} + m_{j})\,\left(  \delta_{kj}-\frac{1}{2}\,\sum_{\ell=e,\mu,\tau}\, U^{*}_{\ell k}\,U_{\ell j}\, \frac{m^{2}_{\ell}}{m^{2}_{W}}\right)\,.
\end{align} 
Here, the constants $e$ and $G_{\text{F}}$ are the elementary charge and the Fermi constant while $U_{\ell k}$ is the unitary mixng matrix,
respectively. As for the masses, $m_{k}, m_{\ell}$, and $m_{W}$ denote the masses of the neutrinos for $k= 1,2,3$, the charged lepton masses for $\ell = e, \mu, \tau$, and the $W$ gauge boson mass. It can
be seen from Eq.\eqref{eq:mmoment} that the magnetic moment is suppressed by the neutrino
mass. For $k=j$ one has the diagonal magnetic moment while the transition magnetic moment ($k \neq j$)
is more suppressed due to the square of the mixing matrix dependence. We focus on the diagonal magnetic moment since it is much larger
than the transition magnetic moment. We do not consider Majorana neutrinos since they only have imaginary transition magnetic moment~\cite{Giunti:2024gec}.

To simplify further, we assume that only one of the neutrino species is non-relativistic today. Therefore, one needs to consider the non-relativistic Hamiltonian to probe the CNB-photons interaction. We choose the Dirac representation of the spinor with 
\begin{align}
\label{eq:Dmatrix}
\gamma^{0}=
\begin{pmatrix} 
1& 0 \\ 0 & -1 
\end{pmatrix},
\alpha^{k}= 
\begin{pmatrix} 
0 & \sigma^{k} \\ \sigma^{k} & 0 
\end{pmatrix}\,.
\end{align}
Taking into account the Dirac mass term of the neutrino, the effective Hamiltonian in the non-relativistic limit becomes~\cite{Aharonov:1984xb} 
\begin{align}
\label{eq:HNR}
H_{\text{NR}} = \frac{1}{2\,m_{\nu}}\, \vec{\sigma}.\left(\vec{p}_{\nu}-i\, \mu^{\text{D}}\,\frac{\vec{E}}{c^{2}}\right)\,\, \vec{\sigma}.\left(\vec{p}_{\nu}+i\, \mu^{\text{D}}\,\frac{\vec{E}}{c^{2}}\right)\,.
\end{align}
Next, we expand the product of Pauli matrices and define the magnetic moment vector $\vec{\mu}^{\text{D}} = \mu^{\text{D}} \, \vec{\sigma}$ to arrive at 
\begin{align}
\label{eq:HNR2}
H_{\text{NR}} = \frac{1}{2\,m_{\nu}} \,\left(\vec{p}_{\nu}-\,\frac{\vec{E}}{c^{2}} \times \vec{\mu}^{\text{D}}\right)^{2} - \frac{(\mu^{\text{D}})^{2}\,E^{2}}{m_{\nu}\,c^{4}}\,. 
\end{align}
where $(\mu^{D})^{2} = \vec{\mu}^{D}.\vec{\mu}^{D}$ and $E^{2} = \vec{E}.\vec{E}$.
This describes the effective Hamiltonian of a neutral particle with non-zero magnetic moment in the
external electric field $\vec{E}$. Since we are dealing with the CNB we need to sum over all CNB neutrinos. To make the notation simpler, we set $m_{\nu} = m_{s}$, $\vec{p}_{\nu} = \vec{p}_{s}$, and $\mu^{\text{D}} = \mu_{s}$ and further sum over the $s$ index to get
\begin{align}
\label{eq:HNRsum}
H_{\text{NR}} =\sum_{s} \left[ \frac{1}{2\,m_{s}} \,\left(\vec{p}_{s}-\,\frac{\vec{E}(\vec{r}_{s})}{c^{2}} \times \vec{\mu}_{s}\right)^{2} - \frac{\mu^{2}_{s}\,E^{2}(\vec{r}_{s})}{m_{s}\,c^{4}}\right]\,, 
\end{align}     
where $\vec{r}_{s}$ is the location of the $s$-th neutrino. Expanding the Hamiltonian to extract the interaction term, we arrive at the following expressions
\begin{align}
\label{eq:Hint0}
H_{I} = H_{I1} + H_{I2}\,,
\end{align}
\begin{align}
\label{eq:Hint1}
H_{I1} = \sum_{s}\left[- \frac{\vec{p}_{s}}{2\,m_{s}}.\left(\frac{\vec{E}(\vec{r}_{s})}{c^{2}} \times \vec{\mu}_{s}\right) -  \left(\frac{\vec{E}(\vec{r}_{s})}{c^{2}} \times \vec{\mu}_{s}\right). \frac{\vec{p}_{s}}{2\,m_{s}}\right] \,,
\end{align} 
\begin{align}
\label{eq:Hint2}
H_{I2} = \sum_{s}\left[ \frac{\left(\vec{E}(\vec{r}_{s})\times \vec{\mu}_{s}\right)^{2}}{2\,m_{s}\,c^{4}}-\frac{\mu^{2}_{s}\,E^{2}(\vec{r}_{s})}{m_{s}\,c^{4}}\right]\,.
\end{align}
For the elastic forward-scattering process relevant to the accumulated photon phase, the $\mu^{2}_{s}\,E^{2}(\vec{r}_{s})$ term contributes at first order in perturbation theory, whereas the $\vec{p}_{s}.(E(\vec{r}_{s})\times \vec{\mu}_{s})$ term contributes through two insertions at second order. In the non-relativistic expansion, their relative contribution at the amplitude level $\mathcal{M}$ scales parametrically as \cite{Chen:2025tlx}
\begin{align}
\frac{\mathcal{M}_{\vec{p}_{s}.(E(\vec{r}_{s})\times \vec{\mu}_{s})}}{\mathcal{M}_{\mu^{2}_{s}\,E^{2}(\vec{r}_{s})}}
\sim
\frac{p_\nu^2}{m_\nu^2}
\ll 1 \, .
\end{align}

Consequently, the contribution generated by the $\vec{p}_{s}.(E(\vec{r}_{s})\times \vec{\mu}_{s})$ term
is suppressed for relic neutrinos satisfying $m_{\nu} \gg p_{\nu}$. Since $\langle p_\nu \rangle = 0.529$ meV, we set the lower bound on the neutrino mass considered here to be 1 meV. 
In this case, the interaction Hamiltonian becomes
\begin{align}
\label{eq:Hintf1}
H_{I}= \sum_{s}\, \frac{E^{2}\,\mu^{2}_{s}}{m_{s}\,c^{4}} \left(\frac{1}{2}\, \text{sin}^{2}\theta_{\mu_{s} E} - 1 \right)\,,
\end{align}
where $\theta_{\mu_{s} E}$ is the angle between $\vec{\mu}_{s}$ and $\vec{E}(\vec{r}_{s})$.
Since we do not know the direction of $\vec{\mu}_{s}$, we assume $\theta_{\mu_{s} E} = 0$.  Actually, one
could rotate the interferometer to get the time modulation of the CNB signal since $\theta_{\mu_{s} E}$
would vary with time. We will consider this scenario in the future work.

Next, we quantize the light to get the quantum description of the light or photons with fixed polarization~\cite{cohen:1987} 
\begin{align}
\label{eq:photon}
\vec{A}(\vec{r}) = \sum_{i} \left[ \frac{\hbar}{2 \,\epsilon_{0}\, \omega_{i} L^{3}} \right]^{1/2} \left[\hat{a}_{i}\, \vec{\varepsilon}_{i} \, e^{\text{i} \vec{k}_{i} \cdot \vec{r}} + \hat{a}^{\dagger}_{i}\, \vec{\varepsilon}_{i} \, e^{-\text{i} \vec{k}_{i} \cdot \vec{r}} \right] \,,
\end{align} 
\begin{align}
\label{eq:Ephoton}
\vec{E}(\vec{r}) = \text{i}\,\sum_{i} \left[ \frac{\hbar\,\omega_{i}}{2 \,\epsilon_{0}\, L^{3}} \right]^{1/2} \left[\hat{a}_{i}\, \vec{\varepsilon}_{i} \, e^{\text{i} \vec{k}_{i} \cdot \vec{r}} - \hat{a}^{\dagger}_{i}\, \vec{\varepsilon}_{i} \, e^{-\text{i} \vec{k}_{i} \cdot \vec{r}} \right] \,.
\end{align} 
Here, we quantize the radiation field with $\hat{a}_{i}$ ($\hat{a}^{\dagger}_{i}$) denoting the annihilation (creation)
operator of the photon field for the i-th mode which satisfies the
commutation relation $[\hat{a}_{i},\hat{a}^{\dagger}_{j} ] = \delta_{ij}$.
We use the box quantization of the radiation field $\vec{A}(\vec{r})$ and $\vec{E}(\vec{r})$ in a volume $L^{3}$, with the associated boundary condition $\vec{k} \cdot \vec{L} = 2\pi\,n$, where $n$ denotes the integer~\cite{Fox:2006quantum}. Note that the relation between the wave number and the angular frequency of the photon is given by $ \omega = |\vec{k}|\, c$.

Following \cite{Nugroho:2024ltb}, we keep only the terms that respect the photon number conservation to arrive at
\begin{align}
\label{eq:HintF}
H_{I} \equiv \hat{H}_{\text{int}} &= -\sum_{s} \frac{\mu^{2}_{s}}{m_{s}\,c^{4}} \, \left[ \frac{\hbar\,\omega}{\epsilon_{0}\, L^{3}} \right] \left(\hat{a}^{\dagger} \hat{a} + \frac{1}{2} \right)\,,\nonumber \\
&= -\frac{\mu^{2}_{\nu}}{m_{\nu}}\,\left[ \frac{\hbar\,\omega^{4}}{8\,\pi^{3} \,\epsilon_{0}\, c^{7}} \right] \left(\hat{a}^{\dagger} \hat{a} + \frac{1}{2} \right)\, N_{\nu}\,, 
\end{align} 
where we assume that all CNB neutrinos have the same magnetic moment $\mu_{s} = \mu_{\nu}$ as well as the same mass $m_{s} = m_{\nu}$. As a result, 
the sum over all neutrinos scales as the total number of CNB neutrinos interacting with
photons $N_{\nu}$. We have also substituted $L = 2\pi/ k$ for $n = 1$ in the second line of Eq. \eqref{eq:HintF} which corresponds to a monochromatic laser. This is due to the fact that a single mode laser is constructed by confining the multimodes
light inside a cavity to form standing waves. The associated wavelengths of these waves are set by the cavity length $L$. A
frequency selective device such as Fabry-Perot etalon is then utilized to select one particular frequency mode to produce a single mode light
field~\cite{Fox:2006quantum}.

If the initial photon state is given by $\ket{\Psi}$, the interaction between photons and the CNB  would change the photon state into $\ket{\Psi^{'}}$. These states are related to each other by unitary transformation~\cite{Kartner:1993,Ou:2017}
\begin{align}
\label{eq:psiP}
\ket{\Psi^{'}} &= \hat{U}_{\delta} \, \ket{\Psi} =e^{-\text{i}\,\hat{H}_{\text{int}}\text{t}/\hbar}\, \ket{\Psi}= e^{\text{i}\,\hat{N}\,\delta_{\text{CNB}}}\,\ket{\Psi}\,, 
\end{align}
where we have used the fact that $ N \equiv \left\langle \hat{a}^{\dagger}\hat{a} \right\rangle\gg 1$ with $ \hat{N} \equiv \hat{a}^{\dagger}\hat{a}$ denoting the photon number operator. Furthermore, from Eqs.~\eqref{eq:HintF} and \eqref{eq:psiP}, one obtains the induced phase shift due to the CNB and photons interaction
\begin{align}
\label{eq:delta}
\delta_{\text{CNB}} = \frac{\mu^{2}_{\nu}}{m_{\nu}}\,\left[ \frac{\omega^{4}}{8\,\pi^{3} \,\epsilon_{0}\, c^{7}} \right] \, N_{\nu}\,t\,.
\end{align}     
We see that the induced phase shift depends on the square of the CNB magnetic moment $\mu_{\nu}$, the total number of CNB neutrino $N_{\nu}$, as well as inversely proportional to the neutrino mass $m_{\nu}$. Moreover, it also depends on the interaction time between the CNB and photons which is given by $\text{L}_{\text{arm}}/c$ where $\text{L}_{\text{arm}}$ is the arm length of the interferometer. To obtain $N_{\nu}$, one needs to count the total number of photons interacting with the CNB along the laser path. For a given CNB number density $n_{\nu}$ and the cross-sectional area of the laser beam $A_{\text{beam}}$ traversing along a trajectory $\ell$ we have
\begin{align}
\label{eq:ell}
N_{\nu} = A_{\text{beam}}\int^{\text{L}_{\text{arm}}}_{0} d\ell\, n_{\nu}\,.
\end{align} 

\section{CNB Induced Phase Measurement Scheme}
\label{sec:phase}
\begin{figure}
	\centering
	\includegraphics[width=0.9\textwidth]{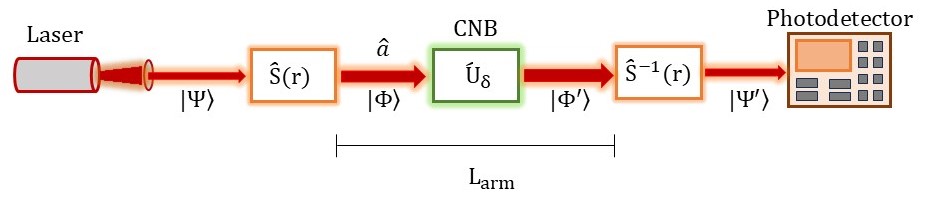}
	\caption{The schematic figure of the single arm interferometer. Here, the photon state changes from $\ket{\Psi}$ to $\ket{\Psi^{'}}$ due to CNB-photon interaction.}
	\label{fig:interfero}
\end{figure}
The induced phase shift from CNB-photon interaction can be probed using laser interferometer. However, it should be noted that the conventional interferometer with two perpendicular arms will not be able to resolve this interaction since both arms experience uniform CNB flux. As a result, the output port of this interferometer would read zero phase shift.     
Fortunately, there is another type of a laser interferometer with single arm that can be utilized to probe the CNB-photon interaction. This type of interferometer was proposed by~\cite{Yurke:1986} and has been widely used in quantum optics experiments  

\begin{figure}
	\centering
	\includegraphics[width=0.95\textwidth]{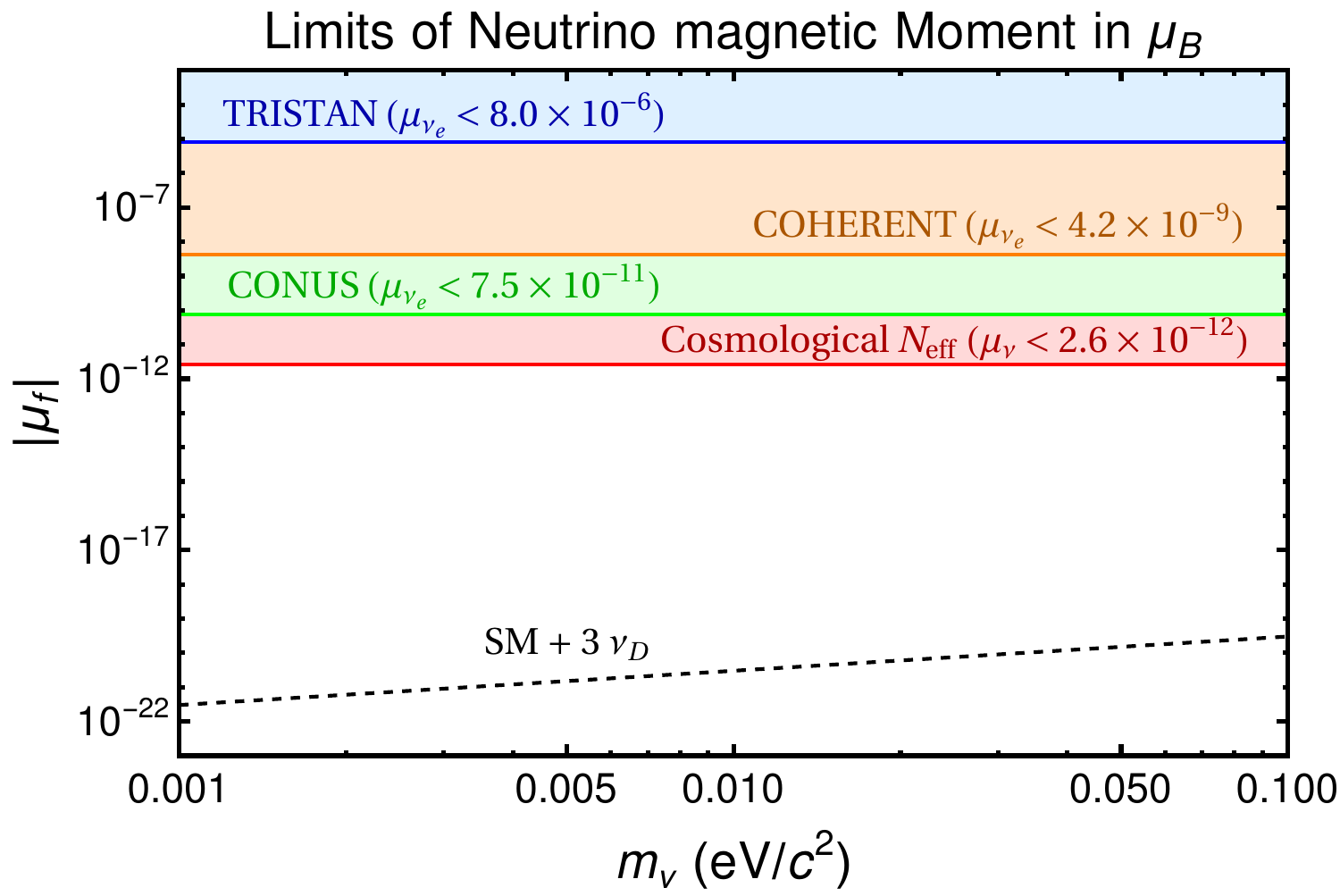}
	\caption{Recent limits on $\mu_{\nu}$ from laboratories and cosmology. The black-dashed line indicates the theoretical value of $\mu^{\text{D}}_{kk}$ given by the SM plus three Dirac neutrinos model.}
	\label{fig:MZInterferometer}
\end{figure}

In Fig. \ref{fig:MZInterferometer}, we show the schematic description of the single arm interferometer with coherent laser source. First, the photon state of the laser source $\ket{\Psi}$ is utilized as an input. Next, the squeezing operator $\hat{S}(r)$ transforms the laser field into the state $\ket{\Phi}$ to become the squeezed coherent state $\ket{\Phi} = \hat{S}(r)\,\ket{\Psi}$. In terms of the vacuum state, the coherent state can be written as~\cite{Ou:2017}
\begin{align}
\label{eq:DefCoherent}
\ket{\Psi}=\ket{\alpha} = e^{-|\alpha|^{2}/2} \sum^{\infty}_{n = 0} \frac{\alpha^{n}}{\sqrt{n !}}\, \ket{n} \equiv \hat{D}(\alpha) \ket{0}\,. 
\end{align} 
Here, $\hat{D}(\alpha) = e^{\alpha\hat{a}^{\dagger}-\alpha^{*} \hat{a}}$ stands for the displacement operator with the corresponding complex displacement parameter $\alpha$.  
The squeezed coherent state is then obained from the vacuum state
\begin{align}
\label{eq:SCoh}
\ket{\Phi}=\ket{r,\alpha} = \hat{S}(r)\,\hat{D}(\alpha)\,\ket{0} =  \hat{S}(r)\,\ket{\Psi}\,.
\end{align}  
Both displacement operator and squeezing operator transform the photon annihilation operator $\hat{a}$ and creation operator $\hat{a}^{\dagger}$ as~\cite{Ou:2017} 
\begin{align}
\label{eq:OpProperties}
\hat{S}^{\dagger}(r)\, \hat{a}\, \hat{S}(r) &= \hat{a}\, \text{cosh}\,r \, + \hat{a}^{\dagger} \, \text{sinh}\,r\,, \\
\hat{S}^{\dagger}(r)\, \hat{a}^{\dagger}\, \hat{S}(r) &= \hat{a}^{\dagger}\, \text{cosh}\,r \, + \hat{a} \, \text{sinh}\,r\,, \\
\hat{D}^{\dagger}(\alpha)\, \hat{a}\,\hat{D}(\alpha) &= \hat{a} \, +\, \alpha\,, \\
\hat{D}^{\dagger}(\alpha)\, \hat{a}^{\dagger}\,\hat{D}(\alpha) &= \hat{a}^{\dagger} \, +\, \alpha^{*}\,,
\end{align}
where $r$ denotes the real squeezing parameter. Subsequently, the photon interacts with the CNB inducing the phase shift $\hat{U}_{\delta}$  to the photon field
\begin{align}
\label{eq: PhiPrime}
\ket{\Phi^{'}} = \hat{U}_{\delta}\,\ket{\Phi}\,.
\end{align}
The anti-squeezing operator $\hat{S}^{-1}(r)$ located at a distance $\text{L}_{\text{arm}}$ from the squeezing opaerator would further transform the photon state into $\ket{\Psi^{'}}$
\begin{align}
\label{eq: psiprime}
\ket{\Psi^{'}} = \hat{S}^{-1}(r) \ket{\Phi^{'}}\,.
\end{align}
In terms of the initial state $\ket{\Psi}$, this can be written as
\begin{align}
\label{eq:totTrans}
\ket{\Psi^{'}} = \hat{S}^{-1}(r)\,\hat{U}_{\delta}\,\hat{S}(r)\,\ket{\Psi} = \hat{S}^{-1}(r)\,e^{\text{i}\,\hat{N}\,\delta_{\text{CNB}}}\,\hat{S}(r)\,\ket{\Psi}\,,
\end{align} 
where we clearly see that in the absence of the CNB-photon interaction $\ket{\Psi^{'}} = \ket{\Psi}$.
At the output port of the interferometer, a photodetector is installed to count the number of photons. This is obtained by evaluating the photon number operator $\hat{a}^{\dagger} \hat{a}$ with respect to the photon final state $\ket{\Psi^{'}}$~\cite{Capolupo:2025zei}
\begin{align}
\label{eq:Nout}
N_{\text{out}}(t) &= \bra{\Psi^{'}} \hat{a}^{\dagger} \hat{a} \ket{\Psi^{'}}\nonumber \\
&= |\alpha|^{2}\left[1- \text{sinh}(2r)\,\text{sin}(2\theta)\, \text{sin}(2\delta(t)) \right]\nonumber\\
&+ |\alpha|^{2} \left[\text{sinh}(4r)\,\text{cos}(2\theta)\,\text{sin}^{2}\delta(t) \right]\nonumber\\
&+ |\alpha|^{2} \left[2\,\text{sinh}^{2}(2r)\,\text{sin}^{2}\delta(t) \right]\nonumber \\
&+ \text{sinh}^{2}(2r)\,\text{sin}^{2}\delta(t)\,,
\end{align}
where the angle $\theta$ is the complex phase of the displacement parameter $\alpha = |\alpha|\,e^{\text{i}\theta}$. Note that in the absence of the phase shift we have $N_{\text{out}} = N_{\text{in}} = |\alpha|^{2}$.  This means that the value of $N_{\text{out}} \neq N_{\text{in}}$
indicates the CNB-photon interaction. Therefore, the signal-to-noise ratio (SNR) depends on the
difference between $N_{\text{out}}$ and $N_{\text{in}}$. Since the photon number fluctuation at the output follows Poissonian distribution\cite{Fox:2006quantum}, the SNR is given by~\cite{Capolupo:2025zei}
\begin{align}
\label{eq:SNR}
\text{SNR} &= \frac{\Delta N(t)}{\sqrt{N_{\text{out}}}}
= \frac{1}{\sqrt{N_{\text{out}}}}\, |\alpha|^{2}\,\text{sin}^{2}\delta(t)\left[\left(\frac{1}{|\alpha|^{2}} + 2 \right)\,\text{sinh}^{2}(2r) + \text{sinh}(4r) \right]\,,
\end{align}
where $\Delta N(t) = N_{\text{out}}(t) - N_{\text{in}}$ and $\delta(t) = \delta_{\text{CNB}}$ is given by Eq.\eqref{eq:delta}. In the following, we set the displacement parameter $\alpha$ to be real which implies $\theta = 0$.  

\section{Projected Sensitivity}
Before proceeding to the discussion about the sensitivity of our proposed setup, let us show the current limits of the neutrino magnetic moment as well as the theoretical value given by Eq.\eqref{eq:mmoment}. For the diagonal magnetic moment, the numerical value is 
\begin{align}
\label{eq:valuemu}
\mu^{\text{D}}_{kk} \simeq \frac{3\,e\,G_{\text{F}}}{8\sqrt{2}\,\pi^{2}}\,m_{k} \simeq 3 \times 10^{-19}\,\left(\frac{m_{k}}{\text{eV}}\right)\,\mu_{\text{B}}\,,
\end{align} 
where $\mu_{\text{B}}$ is the Bohr magneton. This is shown by the black-dashed line in Fig.\ref{fig:MZInterferometer} as a function of neutrino mass $m_{\nu}$. In this plot, we parameterize the magnetic moment as 
\begin{align}
\label{eq:notasimu}
\mu_{\nu} = \mu_{f}\,\mu_{\text{B}}\,,
\end{align} 
where $\mu_{f}$ indicates the dimensionless quantity. Thus, we express the magnetic moment in the unit
of Bohr magneton. We see that the magnetic moment in the simplest extension of the SM is quite small.
This can be compared with the existing bounds from laboratories such as TRISTAN \cite{Tanimoto:2000am}
(blue-solid line), COHERENT \cite{Coloma:2022avw} (orange-solid line), and CONUS \cite{CONUS:2022qbb}
(green-solid line). Although these limits hold for the magnetic moment of the neutrino with electron
flavor, we put it here for comparison with the theoretical value given by the black-dashed line. On the
other hand, the flavor independent limit of $\mu_{\nu}$ comes from the cosmological $N_{\text{eff}}$ \cite{Carenza:2022ngg} (red-solid line) which is still more than eight order magnitudes larger than the theoretical value.

\begin{figure}[t]
	\centering
%	\hfill
	\subfigure{{\includegraphics[width=15.0cm]{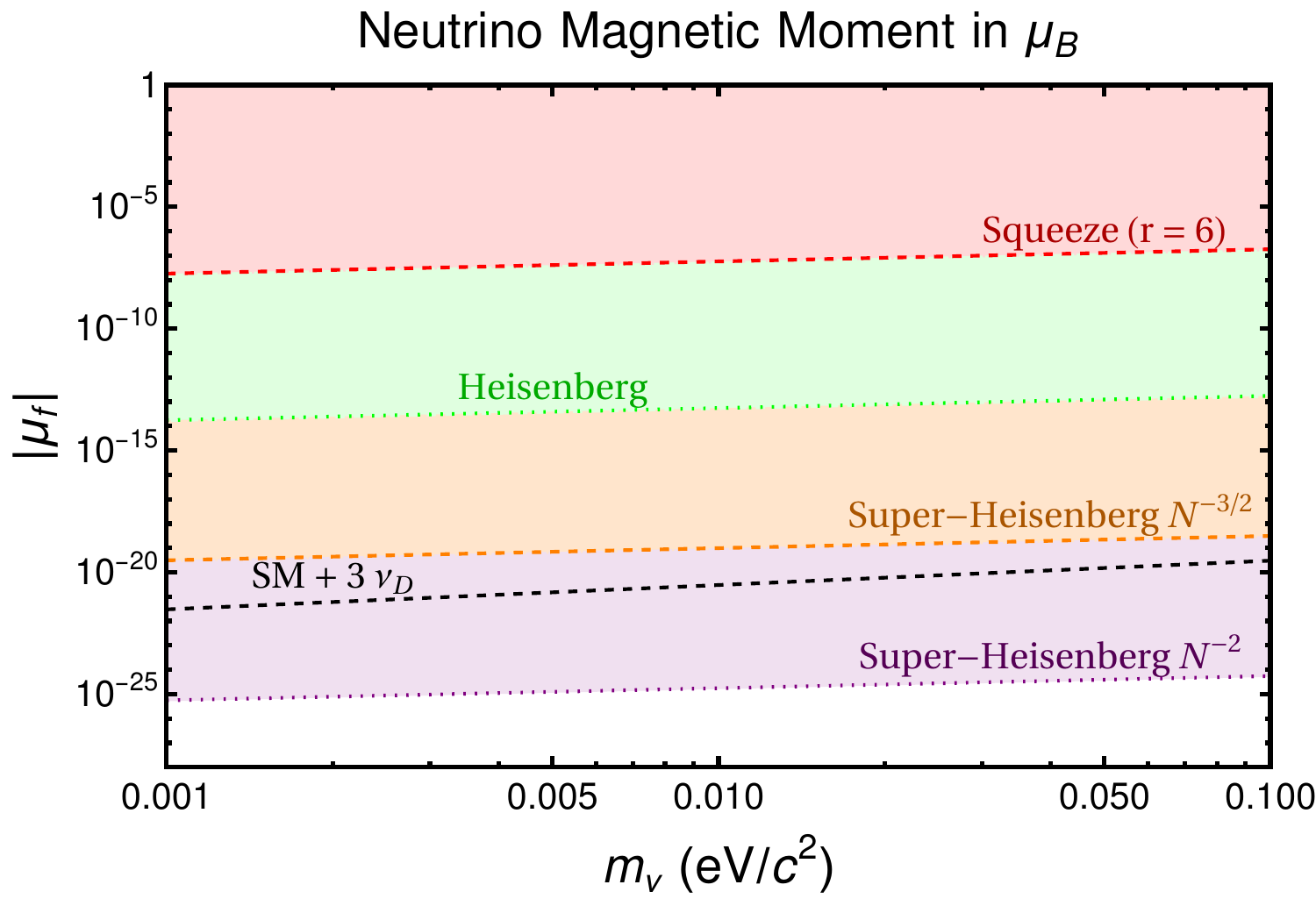} }}
%    \hfill
	\caption{Projected sensitivities of the single arm interferometer with $\text{L}_{\text{arm}} = 1$ km, $A_{\text{beam}}=100 \, \text{cm}^{2}$~\cite{LIGOScientific:2014pky}, and $1.17$ eV laser. Here, we take $n_{\text{CNB}} = 56\, \text{cm}^{-3}$ and the photon number $N = 10^{23}$~\cite{GammeVT-969:2007pci,Bahre:2013ywa,ALPS:2009des,Inada:2013tx}. } 
	\label{fig:sensitivity_astro}
\end{figure}

In Fig.\ref{fig:sensitivity_astro}, we show the sensitivity reach of the single arm interferometer for the neutrino mass range 
$10^{-3}\,\mathrm{eV} \leq m_{\nu} \leq 0.1\,\mathrm{eV}$
allowed by the Particle Data Group (PDG)~\cite{ParticleDataGroup:2024cfk}. This is obtained by setting
the SNR to be equal or greater than one. For the squeezing parameter value $r = 6$, the
projected sensitivity reach is shown by the red-dashed line. This corresponds to the standard quantum
limit (SQL) in the phase measurement. We see that when the interferometer operates at the SQL regime,
it could probe $\mu_{\nu}$ from $10^{-8} \, \mu_{\text{B}} - 10^{-7} \, \mu_{\text{B}}$ in a given neutrino mass range. 
Furthermore, when the interferometer operates at the Heisenberg limit as shown by the green-dotted line, it could
potentially probe the
magnetic moment as small as $10^{-14}\,\mu_{\text{B}}- 10^{-13}\,\mu_{\text{B}}$ surpassing the bound
set by cosmology. A better sensitivity reach is shown by the orange-dashed line in Fig.\ref{fig:sensitivity_astro} which corresponds to the super-Heisenberg limit with $N^{3/2}$ sensitivity enhancement. In this regime, the sensitivity reach of the interferometer could reveal the magnetic
moment within the interval $10^{-19}\,\mu_{\text{B}}- 10^{-18}\,\mu_{\text{B}}$. This is about one
order magnitude larger than the value predicted by the theory (black-dashed line). Finally, the best
projected sensitivity is achieved when the interferometer operates at super-Heisenberg limit with $N^{2}$ sensitivity enhancement as shown by the purple-dotted line. In this regime, the value of $\mu_{\nu}$ as
small as  $10^{-25}\,\mu_{\text{B}}- 10^{-24}\,\mu_{\text{B}}$ could be covered by the single arm
interferometer. This sensitivity reach is three order magitude smaller than the value predicted by the
theory (black-dashed line). Thus, the CNB could potentially be discovered when the interferometer
operates at super-Heisenberg regime with $N^{2}$ enhancement.

\label{sec:result}
\begin{figure}[t]
	\centering
  %  \hfill
    \subfigure{{\includegraphics[width=16.0cm]{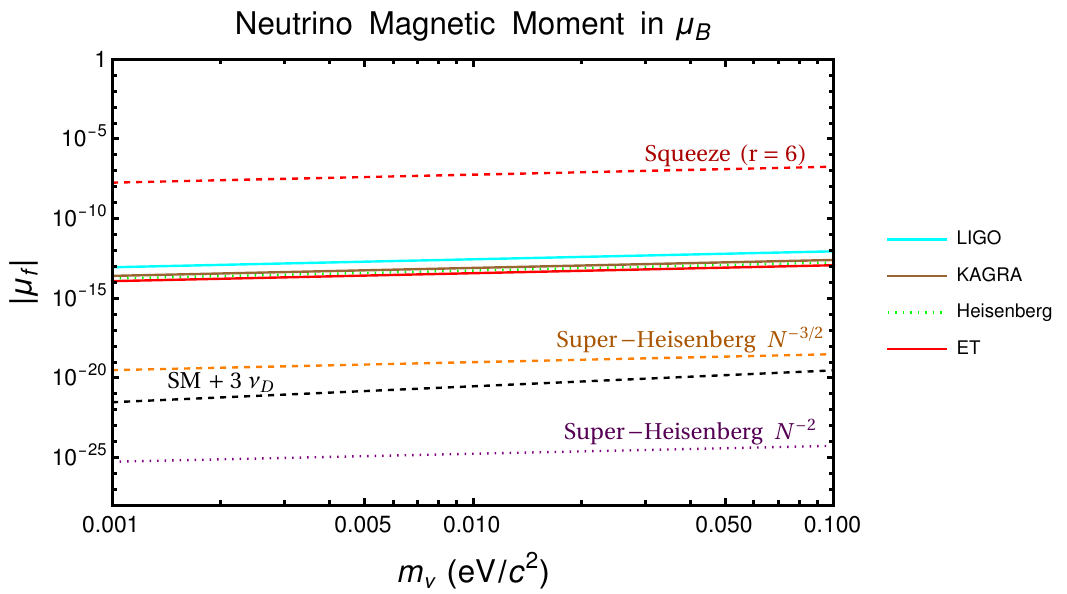} }}
%    \hfill
	\caption{The sensitivity comparison between our setup and GW experiments. Here, we employ 1.17 eV laser and set $v = 1\,\text{km}/\text{hour}$, $A_{\text{beam}}=100 \, \text{cm}^{2}$~\cite{LIGOScientific:2014pky}, and $n_{\text{CNB}} = 56\, \text{cm}^{-3}$. In addition, we fix the photon number $N = 10^{23}$~\cite{GammeVT-969:2007pci,Bahre:2013ywa,ALPS:2009des,Inada:2013tx} and take the value of $t_{1} =$ 1 hour.} 
	\label{fig:sensitivity_labs}
\end{figure}

Next, we would like to compare the sensitivity reach of our proposal with respect to the existing and the
proposed third generation GW detectors. The
sensitivities of GW detectors are constrained by several noise components such as residual gas noise,
gravity gradient noise, thermal noise, quantum noise, seismic noise, etc. The aim of the GW detectors are
to probe the GW signal in a particular frequncy range. In this case, they characterize the noise
components in the detectors by using the power spectral density (PSD). The noise PSD describes the
information about the noise distribution in the particular frequency range. This is to be contrasted with the
constant signal from the CNB. Therefore, to compare the CNB signal in the single arm interferometer
with the sensitivity of GW detectors, we need to model the CNB signal as a function of time

\begin{align}
\label{eq:CNB-noise phase}
\delta(t) = \delta_{\text{CNB}}(t)\, + \delta_{n}(t)\,,
\end{align}       
where $\delta_{n}(t)$ is the phase induced by the noise which in general depends on time. To do this,
we first put the squeezing operator $\hat{S}(r)$ and the anti-squeezing operator $\hat{S}^{-1}(r)$ close to
each other such that the $\text{L}_{\text{arm}}$ equals to zero. Next, we move the anti-squeezing operator $\hat{S}^{-1}(r)$ with constant velocity $v$ during the time interval $0 < t < t_{1}$. The total number of neutrinos interacting with photons during this time interval is
\begin{align}
\label{eq:NnuT}
N_{\nu}(t) = A_{\text{beam}}\, v\,n_{\nu}\,t\,,
\end{align} 
where the arm length now depends on time as $\text{L}_{\text{arm}}(t) = v\,t$. As a result, the induced phase shift due to CNB-photon interaction can be written as 
\begin{align}
\label{eq:phaseT}
\delta_{\text{CNB}}(t) = \frac{\mu^{2}_{\nu}}{m_{\nu}} \left[\frac{\omega^{4}}{8\,\pi^{3}\,\epsilon_{0}\,c^{7}} \right]\,A_{\text{beam}}\, v\,n_{\nu}\,t\, \left(\frac{v\,t}{c}\right)\,,  
\end{align}
where the last term in the parenthesis comes from the interaction time between the CNB and photons $\text{L}_{\text{arm}}(t)/c$.

To obtain the CNB signal in the frequency domain, we do the Fourier transform on $\delta_{\text{CNB}}(t)$ 
\begin{align}
\label{eq:fourier}
\tilde{\delta}_{\text{CNB}}(f) &= \frac{1}{2\pi}\,\int^{\infty}_{-\infty} \,dt\, \delta_{\text{CNB}}(t) \, e^{-\text{i} \,2\pi f t} \nonumber \\
&= \frac{\mu^{2}_{\nu}\,v^{2}}{m_{\nu}} \left[\frac{\omega^{4}}{8\,\pi^{3}\,\epsilon_{0}\,c^{8}} \right]\,A_{\text{beam}}\,n_{\nu}\nonumber \\
&\times \frac{1}{2\,\pi\,(2\pi\,f)^{3}} \left[2\,\text{i} + e^{-\text{i}\,2\pi\,f\,t_{1}}\left(-2\,\text{i} + 2\pi\,f\,t_{1}\left(2+ \text{i}\,2\pi\,f\,t_{1} \right)\right)\right]\,.
\end{align}
\begin{figure}[t]
	\centering
  %  \hfill
    \subfigure{{\includegraphics[width=15.7cm]{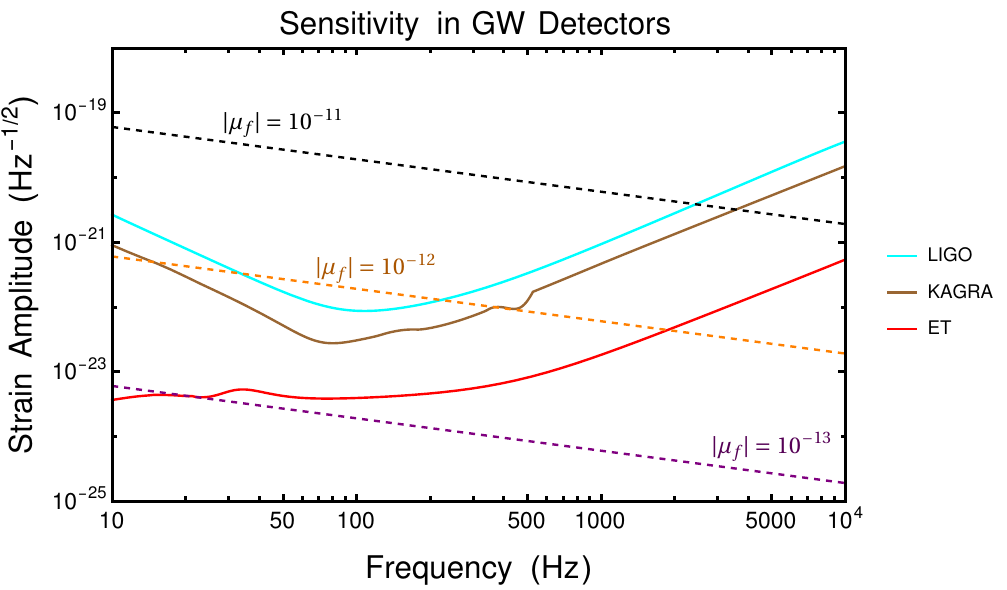} }}
%    \hfill
	\caption{The strain amplitude of LIGO (cyan-solid curve), KAGRA (brown-solid curve), and ET (red-solid curve) taken from \cite{Moore:2019mbg}. The CNB signal correponds to three different values of $\mu_{\nu}$ are also shown for $m_{\nu} =$ 0.05 eV. }
	\label{fig:sensitivity_gw}
\end{figure}
The corresponding  CNB signal power spectral density in the unit of $\text{Hz}^{-1/2}$ is given by\cite{Moore:2014lga}
\begin{align}
\label{eq:sPSD}
\sqrt{S_{\text{CNB}}(f)} = \sqrt{4\,f} \, |\tilde{\delta}_{\text{CNB}}(f)|\,.
\end{align}
Moreover, we extract the PSD of the noise $\delta_{n}(t)$ in a particular GW detector from its
corresponding sensitivity. Finally, the SNR can be obtained by taking the ratio between the squared of
the CNB signal PSD and the squared of the noise PSD integrated over the frequency interval of a given GW
detector \cite{Moore:2014lga}
\begin{align}
\label{eq:gwSNR}
\text{SNR} = \int^{f_{\text{max}}}_{f_{\text{min}}}\, \frac{df}{f} \frac{S_{\text{CNB}}(f)}{S_{n}(f)}\,,
\end{align}
where $S_{n}(f)$ is the squared PSD of the noise in a typical GW detector. To obtain the SNR, we integrate from 10 Hz to $10^{4}$ Hz and further setting the SNR to be greater or equal to one to get the sensitivity line.  
In Fig. \ref{fig:sensitivity_labs} we show the sensitivity of the existing GW experiments such as LIGO  \cite{LIGOScientific:2016aoc} (cyan-solid line), KAGRA \cite{Somiya:2011np} (brown-solid line) as well as the proposed third generation GW
detector, the Einstein telescope (ET) \cite{ET:2020ets} (red-solid line), provided that the two arms of the
interferometers experience different CNB density. We see from this figure that
LIGO and KAGRA sensitivities are close to the Heisenberg limit while  ET is slighlty better than the Heisenberg limit. This shows that LIGO and KAGRA could probe the neutrino magnetic moment as small as $10^{-13}\,\mu_{\text{B}}$ while ET could cover $\mu_{\nu}$ up to $10^{-14}\,\mu_{\text{B}}$.

Furthermore in Fig.\ref{fig:sensitivity_gw}, we show the PSD or the strain amplitude of the CNB signal
in the frequency domain. This is a common language used in GW communities when evaluating noises and
the signal in their detectors. Here, we set $m_{\nu} = $ 0.05 eV and plot three different values of $\mu_{f} : 10^{-11}, 10^{-12},\,\text{and}\, 10^{-13}$ which are shown by the black-dashed line, orange-
dashed line, and purple-dashed line, respectively. The behaviour of the CNB signal in these lines with
respect to the frequency can be understood from the last line of Eq.\eqref{eq:fourier}. The sensitivity
curves from LIGO (cyan-solid line), KAGRA (brown-solid line), and ET (red-solid line)  are also shown.   
We see that the three values of $\mu_{\nu}$ given there could be probed in these GW detectors which is  consistent with our previous results shown in Fig.\ref{fig:sensitivity_labs}.

\section{Summary and Conclusion}
\label{sec:Summary}

The detection of the cosmic neutrino background, produced in early universe, remains as an open problem
in modern physics. Despite of its abundance, it has been a difficult task to do thanks to the
elusive nature of the neutrino as well as its minuscule energy. Detecting the CNB not only would improve our understanding of the early universe, but also uncover the nature of the neutrino itself. 

In this work, we propose to utilize single arm interferometer to probe the relic neutrinos. Our setup
consists of a coherent laser source, one squeezing device, one anti-squeezing device whis is separated
at a distance $\text{L}_{\text{arm}}$ from the squeezing device, and a photodetector. We study four
distinct regimes of the interferometer operation: the standard quantum limit (SQL), the Heisenberg
limit, the super-Heisenberg limit with $N^{3/2}$ enhancement, and another super-Heisenberg limit with $N^{2}$ sensitivity enhancement. We show that the projected sensitivity reach of the single arm interferometer could probe the neutrino magnetic moment as small as $10^{-8}\,\mu_{\text{B}}$, $10^{-14}\,\mu_{\text{B}}$, $10^{-19}\,\mu_{\text{B}}$, and $10^{-25}\,\mu_{\text{B}}$ provided it
operates under the SQL, the Heisenberg limit, super-Heisenberg limit with $N^{3/2}$ and $N^{2}$ sensitivity
enhancement, respectively. We also show that the $N^{2}$ enhancement of the super-Heisenberg regime  could potentially detect the CNB.

We then compare the sensitivity of our setup with respect to the existing  as well as the proposed GW
detectors. We show that LIGO and KAGRA are sensitive to probe the neutrino magnetic moment as small as $10^{-13}\,\mu_{\text{B}}$ while the Einstein Telescope could probe one order magnitude better
provided that their two arms experience different CNB density. This shows that GW detectors sensitivity is close to our single arm interferometer operating at the Heisenberg regime. All in all, we demonstrate that single arm interferometer provide a promising venue to detect the relic neutrinos.

\section*{Acknowledgment}
\label{sec:Acknowledgment}
We would like to express our gratitude to the theoretical physics division of the IPB University for their support and help.

\end{document}